\documentstyle[12pt]{article}

\begin{document}

\begin{flushright}
Preprint Oxford:\\
OUTP-93-30P \\
Preprint IFUNAM:\\
FT94-39 Feb/94\\
\end{flushright}

{}~\\\vskip1truein

\begin{center}{\Large\bf
Effect of wavefunction renormalisation in N-flavour QED$_3$ at finite
temperature\\ \vskip0.2in
}\end{center}
\begin{center}
by

{\bf{I.J.R. Aitchison\footnote{Permanent address. Present address until March
1994: Department of Physics, FM-15, University of Washington, Seattle, WA
98195. Address from March 1994 until September 1994: CERN, CH 1211 Geneva 23,
Switzerland. } and M. Klein-Kreisler\footnote{Present address:
Instituto de Fisica, UNAM, Apartado Postal 20-364, 01000 Mexico, DF,
Mexico.}}}\\\vskip0.1in
Department of Physics\\
Theoretical Physics\\
1 Keble Road\\
Oxford\\
OX1 3NP\\

\vskip0.3in
Submitted to: {\em Physical Review D}
\vskip0.4in
\noindent
{\bf{Abstract}}
\end{center}
\vskip0.05in
\setlength{\baselineskip}{0.2in}
A recent study of dynamical chiral symmetry breaking in N-flavour
QED$_3$ at finite temperature is extended to include the effect of
fermion wavefunction renormalisation in the Schwinger-Dyson equations.
The simple ``zero-frequency'' truncation previously used is found to
lead to unphysical results, especially as $T \to 0$.  A modified set of
equations is proposed, whose solutions behave in a way which is
qualitatively similar to the $T=0$ solutions of Pennington et al. [5-8]
who have made extensive studies of the effect of wavefunction
renormalisation in this context, and
who concluded that there was no critical $N_c$ (at T=0) above which
chiral symmetry was restored.  In contrast, we find that our modified
equations predict a critical $N_c$ at $T \not= 0$, and an $N-T$ phase
diagram very similar to the earlier study neglecting wavefunction
renormalisation.  The reason for the difference is traced to the
different infrared behaviour of the vacuum polarisation at $T=0$ and at
$T \not= 0$.

\newpage

\setlength{\baselineskip}{0.3in}
\noindent
\section{Introduction}\vskip0.1in

In a recent paper with collaborators [1], we studied dynamical chiral
symmetry breaking in $N-$flavour $QED_3$ at finite temperature, in the
large $N$ approximation.  Using an approximate treatment of the
Schwinger-Dyson equation for the fermion self-energy, we found that
chiral symmetry was restored above a certain critical temperature, which
itself depended on $N$.  The ratio $r$ of twice the zero-momentum and
zero-temperature fermion mass to the critical temperature turned out to
have a value of about 10 (approximately independent of $N$), which is
considerably larger than
a typical BCS value, but consistent with previous work [2] using a
momentum-independent self-energy.  We found evidence for a
temperature-dependent critical number of flavours, above which chiral
symmetry was restored, and the $N-T$ phase diagram for spontaneous mass
generation in the theory was presented.  A fuller account of some of the
relevant details, and of the possible relevance to high $T_c$
superconductivity, is contained in [3].

The question of the existence, or not, of a critical $N$ in analogous
calculations at zero temperature is still to some extent controversial.
The original calculations of Appelquist and co-workers [4] found chiral
symmetry breaking only for $N < N_c$ where $N_c = 32/\pi^2$, but this
work made a possibly crucial appeal to perturbation theory (in $1/N$) to
justify the neglect of wavefunction renormalisation and the use of a
simple bare vertex.  This step has been strongly criticized by
Pennington et al. [5,6,7,8], who in a series of papers, using
Schwinger-Dyson equations with
increasingly elaborate non-perturbative vertices satisfying the Ward and
Ward-Takahashi identities, have found no evidence for any $N_c$ -
rather, the fermion mass simply decreases exponentially with $N$.  On
the other hand, an alternative non-perturbative study by Atkinson et al.
[9] suggested that chiral symmetry is restored at large enough $N$; but
this paper is also criticized in recent work of Curtis et al. [8].
Finally, Kondo and Nakatani [10] have examined the effect of
imposing an infra-red cut-off on the Schwinger-Dyson equations,
including wavefunction renormalisation and various ansatzes for
the vertex. For the Pennington-Webb [5] vertex, which we also shall
use, Kondo and Nakatani [10] obtained a (cut-off dependent)  $N_c$ for
the case of a ``large'' infra-red cut-off, and $N_c \to \infty$ for the
case of a ``small'' cut-off.

Our previously mentioned calculations at finite temperature [1] made
precisely the same perturbative approximation as Appelquist et al. [4],
by neglecting wavefunction renormalisation and using the bare vertex.
The arguments and results of Pennington et al. [5-8] certainly provide
strong motivation to go beyond that approximation, and investigate
whether the conclusions of our finite temperature study survive a better
treatment of the vertex.  The present paper reports the results of such
an investigation, using a simplified (but not perturbative) form of
the vertex, but
otherwise following [1] as closely as possible.  In brief, we find that
the conclusions of [1] are, in fact, essentially unchanged, though the
formulation of a simple extension of the formalism of [1] to include
wavefunction renormalisation turns out to be not completely
straightforward. It seems that the crucial ingredient in obtaining
a finite $N_c$ is the
softening of the infra-red behaviour at finite temperature. Our
calculations therefore suggest a natural physical interpretation
of the ``large'' infra-red cut-off regime introduced phenomenologically
(at $T=0$) by Kondo and Nakatani [10].

\vskip0.1in
\noindent
\section{Simple Schwinger-Dyson equations at $T \not=
0$}\vskip0.1in

We shall choose a simplified ansatz for the vertex, and we begin by
introducing that choice within the context of the zero-temperature S-D
equations; then we shall pass to the finite temperature case.  In
Euclidean space, the S-D equation for the fermion propagator is\\
\begin{equation}
S_F^{-1}(p) = S_F^{(0)-1}(p) - e^2 \int \frac{d^3k}{(2 \pi)^3}
\gamma^\mu S_F(k) \Delta_{\mu \nu} (q) \Gamma^\nu(k,p)
\end{equation}
where the superscript 0 denotes the bare quantity, $S_F^{-1}(p)=(1+A(p))
p \hskip-0.07in{/} + \Sigma (p)$ is the inverse fermion propagator, and
$q=k-p$.  We shall not make any change here in our previous (and common)
choice for $\Delta_{\mu\nu}$ - namely, it is approximated by the sum of
massless fermion bubbles.  We shall also continue to work in Landau
gauge.  Turning then to $\Gamma^\nu$, the Ward-Takahashi identity may be
written as
\begin{equation}
q_\nu \Gamma^\nu (k,p) = S_F^{-1}(k) - S_F^{-1}(p).
\end{equation}

\noindent Taking the limit $q_\nu \to 0$ yields the Ward identity
\begin{equation}
\frac{\partial S_F^{-1}(p)}{\partial p_\nu} = \Gamma^\nu(p,p)
\end{equation}
which ensures that the full vertex is free of kinematic singularities.
Both (2) and (3) should hold to all orders of perturbation theory.

Ball and Chiu [11] have given a vertex which satisfies both these
relations, but has an unconstrained transverse part (which, however, is
believed to be unimportant in the infrared region [11, 5-7]).  In fact,
a more general vertex including not only the full Ball-Chiu vertex but
also a non-trivial transverse part has recently been studied by Curtis
et al. [8], at zero temperature.  This vertex satisfies (2) and (3) and
also correctly reproduces the leading asymptotic behaviour known from
perturbation theory.  For our purposes, however, the salient fact is
that the results (as
to dynamical mass generation, $N_c$ etc.) of this most recent work are
in qualitative agreement with earlier studies [5,6] where a much simpler
choice of vertex was made (albeit one no longer satisfying (2) and (3)),
namely
\begin{equation}
\Gamma^\nu(k,p) = \gamma^\nu(1+A(k)).
\end{equation}

Similar approximations in four dimensions have been shown to give
results which are in fairly good agreement with those obtained using the
full vertex.  It seems reasonable to hope that such agreement will
persist in our finite temperature case, and we shall therefore now adopt
the Pennington-Webb [5] vertex (4).

Inserting (4) into (1) and separating the scalar and spinor parts we
find (see also Eqns. (2.35) and (2.36) of [3])

\begin{equation}
{\cal{M}}(p) = \frac{\alpha}{4 \pi^3N}~~~ \frac{1}{1+A(p)} ~~\int d^3k
{}~~\frac{1}{q^2+ \Pi(q)}~~~ \frac{{\cal{M}}(k)}{k^2+{\cal{M}}^2(k)}
\end{equation}
and
\begin{equation}
A(p) = \frac{\alpha}{16 \pi^3Np^2}~~ \int d^3k~~
\frac{(p^2-k^2)^2-(q^2)^2}{q^2[q^2 + \Pi(q)]}
{}~\frac{1}{k^2+{\cal{M}}^2(k)}
\end{equation}
where $\alpha \equiv e^2N$ is understood to be fixed as $N$ varies,
$\Pi$ is the vacuum polarization, and we have introduced the mass
function
\begin{equation}
{\cal{M}}(p) \equiv \frac{\Sigma(p)}{1+A(p)}.
\end{equation}

The explicit appearance of the $1/N$ factor in (6) is, of course, the
reason for the claim that to leading order in $1/N$ wavefunction
renormalisation can be ignored and $A$ set to zero.  But Pennington and
co-workers have convincingly argued [5-8] that, at least at the low
momenta relevant to dynamical mass generation, the integral can provide
a compensating $N$-dependence (as after all happens for $\Sigma$ or
${\cal{M}}$), so that $A$ is by no means of order $1/N$.  These authors
[5] have obtained numerical solutions to equations equivalent to (5)
and (6), which show no sign
of a critical $N_c$.  We follow them in retaining (6), and pass now to
the finite temperature case, following the usual prescriptions.  As in
our previous work [1] (see also [3]) we define
\begin{equation}
\begin{tabular}{lllllll}
$p$ & = & ($P_0$,~\b{p}), & $P = |\b{p}|$, & $p_0 = (2m+1)\pi/\beta$~
{}~~ $(\beta=1/k_BT)$\\
$k$ & = & $(k_0,~\underline{k})$, & $K = |\underline{k}|$, & $k_0 =
(2n+1)\pi/\beta$\\
$q$ & = & $(q_0,~\b{q}),$ & $Q = |\b{q}| = |\b k - \b{p}|$,  & $q_0 =
2(n-m)\pi/\beta$.\\
\end{tabular}
\end{equation}

Integrals over the temporal component of a fermion loop momentum are
replaced by infinite sums over odd Matsubara frequencies, while bosonic
loops are evaluated by summing over even frequences.  The vacuum
polarisation and the fermion functions $A$ and $\Sigma$ become functions
of (the modulus of) the momentum, and of the temperature, and acquire a
discrete index $n$ corresponding to the Matsubara temporal component.
Since we are here concerned with the effect of introducing the
non-perturbative vertex (4), we shall follow [1] and [3] in retaining
only the $\mu = \nu = 0$
component of the photon propagator $\Delta_{\mu\nu}$, and ignore all but
the zero-frequency ($n=0$) component, so that
\begin{equation}
\Delta_{\mu\nu}(q_0,Q,\beta)~ = ~\delta_{\mu 0}~ \delta_{\nu 0}/[Q^2 +
\Pi_0 (Q,\beta)],
\end{equation}
where, to an excellent approximation [1],
\begin{equation}
\Pi_0(Q,\beta)~ = ~\frac{\alpha}{8\beta}~ ~\left[ Q\beta ~+~ \frac{16\ell
n2}{\pi} ~\exp ~\left(\frac{-\pi}{16 \ell n 2}~Q\beta \right) \right] .
\end{equation}

In the same spirit, we shall also ignore the frequency dependence of
the kinematic factors in (6).  In this zero-frequency limit, then, both
$A$ and $\Sigma$ become independent of the Matsubara frequency index,
and the sums over these indices in (5) and (6) can be performed
explicitly so as to yield the following simple equations for the
temperature-dependent mass function ${\cal{M}}(P, \beta)$ and
wavefunction renormalisation $A(P, \beta)$:
\begin{equation}
{\cal{M}}(P, \beta) = \frac{\alpha}{8N\pi^2} ~ \frac{1}{1+A(P, \beta)}
\int d^2 \b{k}~ \frac{{\cal{M}}(K,\beta)}{Q^2 + \Pi_0(Q,\beta)} ~
\frac{\tanh {\beta \over{2}} \sqrt{K^2 +
{\cal{M}}^2(K,\beta)}}{\sqrt{K^2 + {\cal{M}}^2(K,\beta)}}
\end{equation}
and
\begin{equation}
A(P,\beta) = \frac{\alpha}{16N\pi^2} ~ \int d^2\b{k} ~
\frac{(P^2-K^2)^2-Q^4}{P^2Q^2[Q^2 + \Pi_0 (Q,\beta)]} ~ \frac{\tanh {\beta
\over{2}} \sqrt{K^2+{\cal{M}}^2(K,\beta)}}{\sqrt{K^2 + {\cal{M}}^2
(K,\beta)}}.
\end{equation}

With regard to equations (11) and (12), we note that the first (for
${\cal{M}}$) is the same as that for $\Sigma$ in ref. [1], except for the
coefficient in front of the integral which acquires a factor
($1+A(P,\beta))^{-1}$ in the present case.  In [1], of course, $A$ was
set to zero.  We also note that $A$ does not appear under the integral
in (12), but is given by a simple integral involving ${\cal{M}}$.  This
latter feature is a consequence of the ansatz (4); in general we would
have had to deal with two coupled integral equations for ${\cal{M}}$ and $A$.

It would seem that all that now remains is to solve equations (11) and
(12) numerically.  However, it turns out that there is an unsuspected
problem (or so we regard it) with Eq.(12), as we now explain.
\vskip0.05in

\noindent
\section{Problem with the sign of $A$ in the solution of (11) and
(12).}

\vskip0.05in
It is a standard result in zero-temperature field theory (see for
example [12]) that the complete wavefunction renormalisation factor $Z =
1 + A$ satisfies the relation
\begin{equation}
0 \leq Z < 1,
\end{equation}
which implies
\begin{equation}
-1 \leq A < 0.
\end{equation}

Indeed, the angular integral in the zero-temperature expression (6) for
$A$ was evaluated analytically in [5] and found to be negative definite
(which clearly implies $A<0$), and the corresponding numerically
evaluated $Z$ satisfied (13).  Unfortunately, the same results do not
hold for our approximate expression (12) for $A$ at finite temperature,
as we shall now see.

Consider the angular integral in (12), namely
\begin{equation}
I(P,K,\beta) = \frac{K}{\alpha} \int^{2\pi}_0 ~d \phi ~
\frac{(P^2-K^2)^2 - Q^4}{P^2Q^2[Q^2+\Pi_0(Q,\beta)]}
\end{equation}
where $Q^2 = P^2 + K^2 - 2PK \cos \phi$, and we have included the
(dimensionless) factor $K/\alpha$ coming from the two-dimensional phase
space $d^2\b{k}$.  As it is not possible to evaluate $I$ analytically,
we have had to resort to numerical evaluation.  Clearly it is difficult
to give a complete picture of $I$ as a function of all three variables
$P,K$ and $\beta$, but it turns out that for wide ranges of these
variables $I$ is predominantly \underline{positive}, although it is
negative in the region around $K \approx P$.  These features are
illustrated in Figure 1, which
shows $I$ versus $K/\alpha$ for two values of $P$ (one ``large''
compared to the natural scale $\alpha$, the other small), and two values
of (inverse) temperature $\beta$.  It is clear from (15) that $I$ is
always negative at the point $K=P$, and that the quantity $\Pi_0$ acts
as a kind of regulator for the $1/Q^2$ singularity associated with the
photon propagator.  In particular, from Fig. 1 we see that the width of
the region where $I$
 is negative decreases with increasing $\beta$ (decreasing $T$).  These
latter details are understandable from the form of $\Pi_0$, Eqn.(10),
which has a temperature-independent linear term ($\alpha Q/8$) and
another which depends on temperature as $(\alpha C/8 \beta) exp (- \beta
Q/C)$ where $C = (16 \ell n2)/\pi$.

It is of course difficult to read off from Fig.1 what the eventual sign
of $A$ in (12) will be, since much might depend on the relative
weighting attached to the region $K \approx P$, and to the remainder, in
the $K$-integration of (12).  Nevertheless, it does seem likely that,
especially at the low temperatures characteristic of dynamical mass
generation, the relative insignificance of the negative parts of $I$ in
Fig.1 will imply that $A$ in (12) turns out to be positive,
contradicting (14).  This is indeed the case.

We have solved (11) and (12) by an iterative procedure, as follows.  In
zeroth order, we took $A^{(0)} = 0$ and ${\cal{M}}^{(0)} = \Sigma $, the
solution of Eq.8 of our previous paper [1] (which is the same as (11)
with $A=0)$.  Inserting ${\cal{M}}^{(0)}$ into the right hand side of
(12) gave the first iterate $A^{(1)}$.  This was then substituted into
the coefficient in front of the integral in (11), and the latter
evaluated using ${\cal{M}}^{(0)}$ as the input function, so as to yield
the first iterate
 ${\cal{M}}^{(1)}$.  This was substituted back into (12) to give
$A^{(2)}$, and so on.  The procedure was continued until convergence to
within 2 \% was achieved.  Note that, as in [1], we work with a momentum
cut-off at $\Lambda = \alpha$, and scale all momenta, temperatures and
masses by $\alpha$.

For the sake of illustration we consider the case $N=1$.  We have
obtained stable and converged solutions to Eqns (11) and (12), with
properties we now describe.  First, the behaviour of ${\cal{M}}
(P,\beta)$ is qualitatively similar to that of $\Sigma (P, \beta)$ found
in [1] - namely, ${\cal{M}}$ is constant for $P/\alpha\,
{\ \lower-1.2pt\vbox{\hbox{\rlap{$<$}\lower5pt\vbox{\hbox{$\sim$}}}}\ }
10^{-2}$, and falls rapidly to zero for larger
$P/\alpha$.  Further, the zero-momentum value of ${\cal{M}}$ starts to
fall rapidly when $\beta \alpha$ goes from 2000 to 1000, indicating the
possibility of a finite critical
temperature.  Indeed a plot of ${\cal{M}}(0,T)/\alpha$ versus
$k_BT/\alpha$ suggests a critical temperature of order $k_BT_c \sim 10^{-3}
\alpha$, quite similar to the $T_c$'s found in [1].  However, all these
results involve an $A$ which is greater than zero, and hence a $Z$
violating (13).  Fig. 2 shows the corresponding solution of (12) for
$N=1$ as a function of momentum, for different temperatures.  The
behaviour of the full $Z = 1+A$ is
 quite different from that found by Pennington and Webb [5] from the
zero-temperature equation (6) - which, as mentioned above, always
satisfies (14).

We might wonder whether for very low temperatures our solution to (12)
goes negative, but this does not happen.  Instead, as the temperature
decreases the value of $A(0,\beta)$ rises, approaching a value of
approximately 0.6.  This behaviour bears no resemblance to the
zero-temperature result of Pennington and Webb [5].

It is certainly possible that the condition (14), which relies on
unitarity in Minkowski space, may not necessarily hold in the Euclidean
space appropriate to $T \not= 0$.  There may, for example, be
``heat-bath'' creation processes for temperatures above the
pair-creation threshold which would cause a violation of (14).  But we
shall take the view here that the regime relevant to dynamical mass
generation is definitely a low temperature one ($k_BT \ll \alpha)$, and
that consequently we should hope to find an
 $A$ which is negative and qualitatively similar to that of [5].  After
all, the essential aim of the present work is to examine the effect of a
reasonable temperature-dependent extension of [5] on the existence or
otherwise of a critical $N_c$.  We therefore reject the solutions of
(11) and (12) described above, and seek a modified Eqn (12) which will
give a temperature-dependent $A$ satisfying (14), for the (low)
temperature with which we are concerned.\\
\vskip0.05in
\noindent

\section{Modification of the ``$A$'' equation to secure \newline
$-1\!\leq\!A\!<\!0$ at $T\!\not=\!0$.}

We want to understand why the zero-temperature expression (6) for $A$
satisfies (14), while our approximate finite-temperature version of it
gives $A>0$.  We believe the answer lies in our dropping all but the
zero frequency components in $\Pi, \Sigma$, and $A$.  First, it is clear
that if all components are kept, the finite temperature equations must
correctly reduce to the zero-temperature ones as $T \to 0$.  More
particularly, by retaining only the zero component we have effectively
lost a dimension (compare (12) with (6)), and this is crucial for the following
reason.
Suppose we consider the zero temperature limit of (15), in which $\Pi_0
\to \alpha Q/8$.  In this case the integrand of (15) is dominated by
very large positive values associated with the regions close to $\phi =
0$ and $2\pi$, whereas for intermediate values of $\phi$ the integrand
is negative but very much smaller.  This is why, as noted earlier, $I$
is predominantly positive.  However, if we were simply to multiply the
integrand of (15) by $\sin \phi$, so as to mimic the three-dimensional
phase-space at
zero temperature, we would effectively eliminate the unwanted large
positive contribution, and enhance the negative ones.  In fact, we have
checked that introducing such a factor by hand in (15), and integrating
$\phi$ from $0$ to $\pi$, renders $I$ negative for all $P$ and $K$ - and
hence ensures $A < 0$.  Hence we believe that a proper
``reconstruction'' of the full phase space, at least near $T=0$, would
succeed in changing the sign of $A$ as desired.

Unfortunately, it seems a formidable task to attack the fully coupled
equations, including all frequency components.  Instead, we shall seek
here a simple modification of (15), which behaves qualitatively in as
similar a way as possible to the zero-temperature kernel in (6), but is
temperature-dependent.  In this way we hope at least to model the effect
which the Pennington-Webb vertex (4) would have in a more realistic
frequency-dependent calculation.

The modification we propose is motivated in part by the notion that the
approximation of retaining only the zero frequency components is best
justified at high temperatures.  In this (small $\beta$) limit, the
quantity $\Pi_0$ in (10) reduces to the (temperature-dependent) value
($2 \alpha \ell n2/\pi\beta$) - but of course we are really interested in
low temperatures.  However, dynamical mass generation is a low-momentum
phenomenon, and furthermore the integrands in (5) or (6) are certainly
enhanced for $q \approx 0$, so that we may hope that it may be reasonable to
try a
``small $\beta Q$'' approximation, rather than simply a ``small
$\beta$'' one.  In the small $\beta Q$ limit, $\Pi_0$ of (10) reduces
again to ($2 \alpha \ell n2/\pi \beta$), plus ${\cal{O}}(\beta Q)^2$
corrections.  These considerations lead us to explore the result of
replacing $\Pi_0$ in (12) by a constant, $\Delta^2$ say, which we will
take to be an adjustable parameter with a value of order $\alpha^2$ or
less.  We will choose it so as to obtain a wavefunction renormalisation
as much like ref. [5] as possible.

In replacing $\Pi_0$ in (12) by $\Delta^2$ we are of course also altering the
$Q \to 0$ behaviour of the kernel, making it less singular. To  ``compensate''
for this we might think of dropping the $K$ in the phase-space factor, at the
same time as replacing $\Pi_0$ by $\Delta^2$. To explore these possibilities
we shall study (a) the ``exact'' angular integral (with the $K$ factor
included) of (15), called $I(P,K,\beta)$; (b) the modified kernel with
$\Pi_0 \to \Delta^2$ and no $K$-factor, namely
\begin{equation}
I_\Delta (P,K) = \int^{2\pi}_0 \frac{(P^2-K^2)^2-
Q^4}{P^2Q^2 (Q^2+ \Delta^2)}d\phi;
\end{equation}
and (c) the kernel with $\Pi_0 \to \Delta^2$ but retaining the $K$-factor,
i.e. the quantity $K\times I_\Delta$. Of these we shall prefer the one which,
on
the one hand, best captures the low-momentum behaviour of $I(P,K,\beta)$
(since this is the relevant region for dynamical mass generation),
and on the other gives a wavefunction which most resembles the
qualitative behaviour of the $T=0$ results of ref. [5].

It turns out that $I_\Delta$ can be integrated analytically, with
the result
\begin{equation}
I_\Delta = \frac{-2\pi}{P^2} \left[ 1 - \frac{|P^2-K^2|}{\Delta^2} +
\frac{[(P^2-K^2)+\Delta^2][(P^2-K^2)-\Delta^2]}{\Delta^2 \{[(P-K)^2 +
\Delta^2][(P+K)^2 + \Delta^2]\}^{{1\over{2}}}} \right].
\end{equation}
Figure 3 shows a plot of (17) as a function of $P$ and $K$ (scaled by
$\alpha$),
for the case $\Delta^2=\alpha^2$.  We note that, most importantly, $I_\Delta$
is almost always {\it negative} (in contrast to $I$ of (15)) except for a small
region of large $K$ and $P$ values, which is not the important region for
dynamical mass generation. In fact, the behaviour of $I_\Delta$ is
similar to that of an analogous integral which arose in an approximate
treatment [6] of the three-dimensional zero-temperature equation (6):
namely, a rapid variation with $K$ for $K<P$, and a much slower variation
for $K>P$, the sign being (in [6]) negative throughout. These features are
true of $I_\Delta$ for a substantial range of $\Delta^2$, and ensure
$A<0$ as we shall see in the next section.

Figure 4 shows a comparison of $I$, $I_\Delta$ and $K\times I_\Delta$ for
$P/\alpha=6.851\times 10^{-4}$ and $\Delta^2=\alpha^2$. It is remarkable how
well the low
$K$ behaviour of $I$ is captured by $I_\Delta$, and how poorly $K\times
I_\Delta$
performs. The large positive contribution in $I$ for larger values of
$K$ is, as noted above, an undesirable feature and the reason for
obtaining $A>0$ in section 3. On the other hand, if too small a value of
$\Delta^2$ is chosen (for example $\Delta^2$ of order $0.1 \alpha^2$) we find
that
$I_\Delta$ becomes too large and negative, with the consequence that
the corresponding $A$ begins to approach -1.  From the appearance of the
factor $(1+A)^{-1}$ in (11) we expect that such $A$'s will be associated with
too large a value of $\cal{M}$, contradicting the basic assumption $\cal{M} \ll
\alpha$.
The above considerations already strongly suggest that $I_\Delta$ will
be the most satisfactory kernel at $T\neq 0$, and we now proceed to
discuss the results of using this kernel in place of $I$ in (12); we
shall also briefly describe the results of using $K\times I_\Delta$.

\section{Results using a modified ``$A$'' equation}

We have solved the original Eqn (11), together with one or the other
of the following modified equations for $A$:
 \begin{equation}
A_\Delta(P,\beta) = \frac{\alpha}{16N\pi^2} \int^\alpha_0 dK I_\Delta(P,K)
\frac{\tanh \frac{\beta}{2}\sqrt{K^2 +
{\cal{M}}^2(K,\beta)}}{\sqrt{K^2+{\cal{M}}^2(K,\beta)}},
\end{equation}

\begin{equation}
A_{K\Delta}(P,\beta) = \frac{\alpha}{16N\pi^2} \int^\alpha_0 dK K I_\Delta(P,K)
\frac{\tanh \frac{\beta}{2}\sqrt{K^2 +
{\cal{M}}^2(K,\beta)}}{\sqrt{K^2+{\cal{M}}^2(K,\beta)}},
\end{equation}
with $I_\Delta$ given by Eqn. (17).  We followed the same iterative procedure
as described in Section 3, obtaining stable and converged solutions
for $A$ and $\Sigma$. The results using $A_{K\Delta}$ can be quickly
summarised. Figure 5 shows $A_{K\Delta}(P,\beta)$ versus $P/\alpha$ for
$\Delta^2=\alpha^2$ and $N=1$; we note that $A_{K\Delta}$ is negative, as
required,
but very small in magnitude. Clearly this is a consequence of the very
small (and negative) value of $K\times I_\Delta$ as seen in Fig. 4.  Figure 6
shows the corresponding ${\cal M}(0,T)/\alpha$ versus $k_BT/\alpha$, for $N=1$.
This
figure is very similar to the $A=0$, $N=1$ case shown in fig.3 of ref. [1], as
would be expected from the small value of $A_{K\Delta}$ shown in Fig.5.
In particular, the ratio $r=2{\cal{M}}(0,0)/k_BT_c$ remains close to 10.
Qualitatively,
the effect of the small negative $I_{K\Delta}$ is to produce a
small upward shift of the ${\cal M}(0,T)/\alpha$ curve, relative to the $A=0$
case; this is easily understood as being associated with the
$(1+A)^{-1}$ factor in (11).  We have obtained the corresponding phase
diagram using $I_{K\Delta}$, but it differs only rather slightly,
and predictably, from fig. 7 of ref. [1].

The results using $I_\Delta$ are more interesting.
For $\Delta^2 \stackrel{<}\sim 0.1 \alpha^2$, we found that $A_\Delta$
was positive for a small range of $P$, near
$P \approx \alpha$; otherwise, $A_\Delta(P)$ was negative as required.  More
seriously, for this value of $\Delta^2$ we found that ${\cal{M}}(P)$ was
only about one order of magnitude smaller than $\alpha$, so that the
condition ${\cal{M}} \ll \alpha$ characteristic of dynamical mass
generation in QED$_3$ (and of
[5]) was no longer true (implying that the approximation of $\Pi$ by the
\underline{massless} bubble would need re-examination).  For $\Delta^2 =
\alpha^2$, however, we did obtain values for  ${\cal{M}}$ and $A_\Delta$ which
were in much closer qualitative agreement with the zero-temperature
results, and we now concentrate on the results for this value of
$\Delta^2$.

In Figure 7 we show ${\cal{M}}(P,\beta)$ as a function of momentum, for
$N=1$ and different $\beta$'s, while Fig. 8 shows the corresponding
$A_\Delta$'s, all satisfying (14).  These $A_\Delta$'s are all very
reasonable-looking,
reminiscent of those at $T=0$ [5,6].
The rapid decrease of ${\cal{M}}$ as
$\beta$ decreases from $\beta = 500/\alpha$ to 130/$\alpha$ suggests the
existence of a critical temperature near $10^{-2}\alpha$.  In Fig. 9 we
show ${\cal{M}}(0,T)$ as a function of temperature, from which it
appears that the critical temperature $T_c$  is such that $k_BT_c \approx
7.8 {\rm{x}} 10^{-3}\alpha$.
Figures 7 and 9 can be compared with Figures 2 and 3 of [1];
the conclusion is that the effect of including wavefunction
renormalisation in the way described is to increase both the mass
${\cal{M}}$ and the critical temperature $T_c$ by roughly a factor of two;
again, this can be qualitatively understood as a consequence
of the $(1+A)^{-1}$ factor in (11).
 Indeed, the ratio $r$ ($ = 2{\cal{M}}(0,0)/k_BT_c$) comes out as
10.27, almost identical to the value for $N=1$ which we quoted in [1]
without using wavefunction renormalisation.

Fig. 8 shows that the dependence of $A_\Delta$ on $\beta$ is not very strong,
disappearing altogether at the high momentum end.  In Fig. 10 we show
$A_\Delta$
at $P=0$, as a function of $T$ (note that (18) actually has non trivial
solutions even for ${\cal{M}} = 0$, so that we can investigate the
region $T>T_c$ for $A_\Delta$).  It can be shown easily that $A_\Delta(P=0, T)
\to 0$
as $T \to \infty$.

Finally, we come to the main question we originally set out to answer -
the existence of a critical $N_c$, or otherwise.  Fig. 11 shows
${\cal{M}}(0, \beta)$ versus $N$, for various fixed temperatures.  As in
our previous work [1], we are not able to obtain reliable results for
${\cal{M}}/\alpha$ much below $10^{-5}$.  Nonetheless, as before, it
seems reasonable to infer that ${\cal{M}}$ does vanish beyond some
finite $N_c$ which itself depends on $T$, increasing as $T$ decreases.
We can extrapolate the ${\cal{M}}/\alpha$ curves to find the critical
values $N_c(T)$ for the various temperatures, and thus obtain the
phase diagram shown in Figure 12, and which is very similar to the one
we obtained previously with $A=0$ [1].

Comparing Figure 11 with Figure 5 of [1], we notice that for a given
temperature, the critical number of flavours has increased in the case
where $A \not= 0$: for example, at $\beta = 10^4/\alpha$, $N_c$ was just
above 1.8 in the $A=0$ case, whereas it now has a value slightly above
2.2.  Although probably too much should not be made of such a relatively
small difference, it is in fact simple to understand it, following an
argument of Pennington and Webb [5].  Comparing our (11) above with (8)
of [1], we can see that a measure of the effect of $A \not= 0$ can be obtained
by
replacing $N$ in the $A = 0$ case by $N(1+\langle A \rangle)$ where
$\langle A \rangle$ is an
average value for $A$.  Hence a solution of the $A=0$ equations for a
given $N_{A=0}$ will effectively correspond to a solution of the $A
\not= 0$ equations with the identification $N_{A\not=0}(1+\langle A \rangle) =
N_{A=0}$.  Since $\langle A \rangle<0$, the critical number of flavours for a
given
$T$ increases in going to the $A \not= 0$ case.

Fig. 13  shows $A_\Delta$ as a function of $P$ for various values of $N$, at
fixed $T$. Although we have not been able to explore as wide a range of
$N$'s as Pennington and Walsh [7], this figure is qualitatively similar
to their Figure 3, for $N \approx 2$, which encourages us to think that
our modified $A_\Delta$-equation, (18), is physically reasonable.  However, it
is interesting to note that in the zero-temperature results of [5] and
[7], $A(P=0)$ seems to approach -1 with increasing $N$, suggesting that
the corresponding $N_{eff} = (N_{A=0})/(1+\langle A \rangle)$ becomes so large
that
any sign of
criticality in $N$ (which might have been present in the $A=0$ case)
disappears.  Even for smaller $N$'s, where $\langle A \rangle$ is considerably
different from -1, the results of [5]-[8] do not show any indication of
a sudden decrease in the dynamical mass at some critical $N$, such as is
seen in our Fig. 11.

Nevertheless, it should be stressed that we are not able to obtain
reliable numerical results for temperatures below $\sim 10^{-4}\alpha$,
so that we are not able to say what the precise zero-temperature limit
of our $N_c(T)$ might be - if indeed it exists at all.  In addition, we
must not forget the inexact nature of the ``$\Delta^2$ modification''.
In short, it is quite possible that $N_c(T) \to \infty$ as $T \to 0$,
which would be in agreement with the conclusion of Pennington et al.
[5-8].

If this suggestion is correct, we need to understand why our non-zero
temperature results still point so clearly to a critical $N$, even
though we have included a non-zero $A$.  The answer seems surely to lie in
the important alteration which the finite-temperature vacuum
polarisation $\Pi_0$ makes to the infrared regime of the fermion
self-energy.  As we have discussed earlier, the $Q \to 0$ limit of
$\Pi_0$ in Eqn. (10) is a finite temperature-dependent term $2\alpha
\ell n2/\pi\beta$ - a phenomenon called
 ``thermal screening''.  The ``$\Delta^2$ modification'' we used in
equation (18) for $A_\Delta$ also captures this softening, for small $Q\beta$.
By contrast, the zero-temperature $\Pi$ of Pennington et al. [5-8]
(inherited from Pisarski [13]) behaves as $\alpha Q$ for small $Q$ and
has no infrared softening.  We believe that it is this infrared
screening, associated with the temperature-dependent $\Pi_0$, which is
crucial for the existence of an $N_c(T)$, just as it was found to be
vital to the generation of a large value for $r$ [1].\vskip0.05in

\noindent
\section{Conclusions}\vskip0.05in

The work of Pennington et al. [5-8] strongly indicated that the
existence [4] of a critical number of flavours $N_c$, above which chiral
symmetry was unbroken in QED$_3$ at zero-temperature, was an artefact of
incorrectly ignoring wavefunction renormalisation (via an unjustified
appeal to perturbation theory in $1/N$).  We have been interested in
extending the study of chiral symmetry breaking in QED$_3$ to finite
temperature.  In [1] we found clear evidence for the existence of an
$N_c$, dependent on $T$,
 but we made the approximation of ignoring wavefunction renormalisation.
 The present study has been aimed at removing that approximation, and
studying the effect of including wavefunction renormalisation on chiral
symmetry breaking in QED$_3$ at $T \not= 0$.

The simplest generalisation of the model of [1] to include wavefunction
renormalisation, in which we adopted the ``zero-frequency''
approximation, turned out to lead to unphysical results, in that they
failed to show any similarity to the zero-temperature results of [5-8]
as $T \to 0$.  We regarded this as reason for discarding that model, and
adopted instead a modified equation for the wavefunction renormalisation
at $T \not= 0$, which gave results consistent with the $T=0$ case.  We
found, using the modified model (namely the kernel
$I_\Delta$ of sections 4 and 5), that although wavefunction
renormalisation was now included
in a way qualitatively very similar to that of [5] - [8], nevertheless
at finite temperature we still found clear evidence for an $N_c$, in contrast
to the results of [5]-[8].  The
essential reason for the difference seemed to be the characteristically
different infrared behaviour of the vacuum polarisation, at zero and at
finite temperature.  Indeed, the inclusion of wavefunction
renormalisation, in the manner described in Section 4 above, gave
results  very little different, overall, from  those of [1]. In particular, a
value of about 10 for the ratio r seems to be remarkably robust, while
the values of ${\cal M}(0,0)$ and $T_c$ depend more on the model for $A$.

The conclusion that it is the infra-red behaviour which is crucial
in obtaining a finite $N_c$ is supported by the calculations of
Kondo and Nakatani [10]. Using (among other vertices) the
Pennington-Webb vertex of equation (4), these authors found
that with an infra-red
cut-off $\epsilon$ of order $10^{-4} \alpha$, a fermion mass was generated for
$N<N_c$, where  $N_c$ depended logarithmically on $\epsilon$. Very roughly, we
might interpret $\epsilon$  as corresponding, in our calculation, to the
``thermal mass'' $(\Pi_0)^{1/2}$, which is of order $T^{1/2}$.  We then have
a physical interpretation of the perhaps somewhat artificial cut-off
introduced by Kondo and Nakatani [10].

 These authors [10] also found that the behaviour of the dynamically
generated mass for $N$ near $N_c$, using the vertex of equation (4), was
consistent with mean field theory: that is, ${\cal M} \approx (N_c-N)^{1/2}$.
By contrast, in our previous paper [1] we presented evidence
(see figure 6 of [1]) which suggested a behaviour of the form
 ${\cal M} \approx  \exp[-C/(N_c-N)^{1/2}]$. However, we have re-examined those
calculations and are now less confident that a firm conclusion can
be drawn regarding the behaviour near $N_c$, where the mass is very
small and numerical accuracy increasingly difficult to control. In fact,
a more conservative interpretation of both our present and previous
data is that they are quite consistent with mean field theory, at least
in the regime (for $N$ not too close to $N_c$) where the calculations
are most reliable.

The present calculation can, of course, be criticised for invoking the
somewhat ad hoc ``$\Delta^2$ modification'' for the $A$ equation (though
we stress again that this modification does the job required of it,
namely to produce sensible looking solutions at low $T$).  It would
clearly be much more satisfactory to attack the equations for the
coupled $n \not= 0$ components of $\Sigma$, using the full $n \not= 0$
components of the photon propagator as given in [3].  In this way the
true effects of the $n \not= 0$ components, especially at low $T$, could be
identified.  No
doubt a start on this problem could be made by going back to the
simplifying approximation $\Sigma(P,\beta) \approx \Sigma(0,\beta)$,
which is likely to be quite reliable (see Eqn (2.54) of [3]).  We hope
to return to this problem elsewhere.\vskip0.05in

\vskip0.1in
\noindent
{\large\bf Acknowledgements}
\vskip0.05in

We are very grateful to Mike Pennington for several illuminating
discussions about the work of references 5-8, and for making available
to us copies of refs. 5 and 6 in particular.  We have also enjoyed
useful discussions with Cesar Fosco, Toyoki Matsuyama and Nick Mavromatos.
The work of I.J.R.A., while at the University of Washington, was supported
in part by a grant from the U.S. Department of Energy to the nuclear
theory group at the University of Washington, by the N.S.F. under grant
No. DMR-9220733, and by the Physics Department, University of Washington.
I.J.R.A. is grateful to the members of the Physics Department, University of
Washington, and especially to Professors David Thouless and Larry Wilets, for
their hospitality and support. M.K.K. wishes to thank the National University
of Mexico, SERC and the sub-department of Theoretical Physics at Oxford for
financial support.
\vskip0.15in

\newpage
\noindent
{\large\bf References}
\vskip0.05in
\begin{enumerate}

\item I.J.R. Aitchison, N. Dorey, M. Klein Kreisler and N.E. Mavromatos,
Phys. Lett. {\bf B294} (1992) 91.

\item N. Dorey and N.E. Mavromatos, Phys. Lett. {\bf B266} (1991)
163.

\item N. Dorey and N.E. Mavromatos, Nucl. Phys. {\bf B386} (1992)
614.

\item T.W. Appelquist, M. Bowick, D. Karabali and L.C.R. Wijewardhana,
Phys. Rev. {\bf D33} (1986) 3704; T.W. Appelquist, D. Nash and
L.C.R. Wijewardhana, Phys. Rev. Lett. {\bf 60} (1988) 2575.

\item M.R. Pennington and S.P. Webb, preprint BNL-40886 (January 1988),
unpublished.

\item D. Atkinson, P.W. Johnson and M.R. Pennington, preprint BNL -
41615 (August 1988), unpublished.

\item M.R. Pennington and D. Walsh, Phys. Lett. {\bf B253} (1991)
246.

\item D.C. Curtis, M.R. Pennington and D. Walsh, Phys. Lett.
{\bf B295} (1992) 313.

\item D. Atkinson, P.W. Johnson and P. Maris, Phys. Rev. {\bf D42}
(1990) 602.

\item K.-I. Kondo and H. Nakatani, Prog. Theor. Phys. {\bf 87} (1992) 193.

\item J.S. Ball and T.W. Chiu, Phys. Rev. {\bf D22} (1980) 2542.

\item C. Itzykson and J.-B. Zuber, {\it Quantum Field Theory},
(McGraw-Hill, Singapore, 1985).

\item R.D. Pisarski, Phys. Rev. {\bf D29} (1984) 2423.

\end{enumerate}
\newpage
\noindent{\large\bf Figure Captions}

\begin{enumerate}

\item[Fig. 1] Angular integral $I(P,K,\beta)$  at $P = 6.85
\times 10^{-4}\alpha, 0.809 \alpha$ and $\beta = 1000/\alpha,
100/\alpha$.

\item[Fig. 2] The wavefunction renormalisation as a function of scaled
momenta, at various (scaled) temperatures, for $N=1$.

\item[Fig. 3] Angular integral $I_\Delta(P, \lambda)$ for $\Delta^2 =
0.3$.

\item[Fig. 4]  Comparison of $I$, $I_\Delta$ and $K\times I_\Delta$ for
$P/\alpha=6.851\times 10^{-4}$ and $\Delta^2=\alpha^2$.

\item[Fig. 5]  $A_{K\Delta}(P,\beta)$ versus $P/\alpha$ for
$\Delta^2=\alpha^2$ and $N=1$.

\item[Fig. 6] The scaled self-energy $m(T)/\alpha$ $(\equiv {\cal
M}(0,T)/\alpha)$ as a function of scaled temperature
for $N=1$ and $\Delta^2 = \alpha^2$ using $A_{K\Delta}$.

\item[Fig. 7] The scaled self-energy as a function of scaled
momentum for $N=1$ and $\beta \alpha = 10^4$, 2000, 500 and
130.  The parameter $\Delta^2$ is fixed at the value $\Delta^2 =
\alpha^2$.

\item[Fig. 8] $A_\Delta(P,\beta)$ as a function of scaled momentum for fixed
$N=1$ and $\beta\alpha = 10^4, 2000, 500$ and 130, with
$\Delta^2=\alpha^2$.

\item[Fig. 9] The scaled self-energy $m(T)/\alpha$ $(\equiv {\cal
M}(0,T)/\alpha)$  as a function of scaled temperature
for $N=1$ and $\Delta^2 = \alpha^2$ using $A_\Delta$.

\item[Fig. 10] $A_\Delta(0,T)$ as a function of scaled temperature for $N=1$
and
$\Delta^2 = \alpha^2$.

\item[Fig. 11] The scaled self-energy $m(\beta)/\alpha$ $(\equiv {\cal
M}(0,\beta)/\alpha)$  as a function of $N$ at various
scaled temperatures.

\item[Fig. 12] The phase diagram of $QED_3$, in our approximation, with
non-trivial wavefunction renormalisation.  The critical line separates
the region where there is dynamical mass generation $({\cal{M}} \not=
0)$ from that in which there is not $({\cal{M}} = 0)$.

\item[Fig. 13] The wavefunction renormalisation $A_\Delta(P,\beta)$ as a
function of scaled momenta, for $N~=~1,~ 1.5, ~2, ~2.2$ at $\beta\alpha =
10^4$.

\end{enumerate}

\end{document}